\documentclass[prb,twocolumn,showpacs,preprintnumbers,amsmath,amssymb,floats]{revtex4}
\usepackage{graphicx}
\usepackage{amsmath}
\usepackage{epstopdf}
\usepackage{amsbsy}
\usepackage{color}
\usepackage[font=small,justification=raggedright]{caption,subfig}
\usepackage{dcolumn}
\usepackage{bm}

\newcommand{\down}{\downarrow}
\newcommand{\up}{\uparrow}
\newcommand{\BSCCO}{{Bi$_2$Sr$_2$CaCu$_2$O$_8$}}

\include{newcommands}

\begin{document}

\title{Local modulations of the spin-fluctuation mediated pairing interaction by impurities in $d$-wave superconductors}
\author{A. T. R\o mer$^1$, S. Graser$^2$, T. S. Nunner$^3$ P. J. Hirschfeld$^4$, and B. M. Andersen$^1$}
\affiliation{$^1$Niels Bohr Institute, University of Copenhagen, Universitetsparken 5, DK-2100 Copenhagen,
Denmark\\
$^2$Theoretical Physics III, Center for Electronic Correlations and Magnetism, Institute
of Physics, University of Augsburg,
D-86135 Augsburg, Germany\\
$^3$Dahlem Center for Complex Quantum Systems and Fachbereich Physik, Freie Universit\"at Berlin, 14195 Berlin, Germany
\\
$^4$Department of Physics, University of Florida, Gainesville, Florida 32611, USA}

\date{\today}

\begin{abstract}

We present a self-consistent real space formulation of spin-fluctuation mediated $d$-wave pairing.  By calculating all relevant inhomogeneous spin and charge susceptibilities in real space within the random phase approximation (RPA), we obtain the effective pairing interaction and study its spatial dependence near both local potential and hopping impurities. A remarkably large enhancement of the pairing interaction may be obtained near the impurity site.  We discuss the relevance of our result to inhomogeneities observed by scanning tunneling spectroscopy on the surface of cuprate superconductors.

\end{abstract}


\maketitle

\section{Introduction}

Scanning tunneling spectroscopy (STS)
experiments on surfaces of several high-$T_c$
materials\cite{Fischer} have discovered that the electronic structure of
at least some of the cuprate superconductors is strongly inhomogeneous in real space.  Modulations observed  include checkerboard local density of states (LDOS)
patterns which peak at biases near the local gap energy,  together with strong gap
modulations  on nanometer length scales.\cite{cren,howald,davisinhom1,davisinhom2,YKohsaka:2007} While the latter modulations appear to be random in character and driven by disorder, the STS conductance maps contain important
correlations of gaps with other physical observables.   For example, the size of the local gap was found to be positively
correlated with  atomic scale defects,
thought to be interstitial oxygen dopants\cite{KMcElroy_05} (see however Ref. \onlinecite{Hoffman}). This observation inspired a reexamination of the
 effect of an impurity in these systems, since it had been anticipated that an oxygen,
which donates two holes to the CuO$_2$ plane, would locally overdope the
system and lead to a smaller gap in the neighborhood of such a dopant impurity. The positive correlation
between the putative oxygens and the observed spectral gap led Nunner {\it et
al.}\cite{nunner:2005,nunner:2007} to propose that the dopants might be
enhancing the pair interaction on the atomic scale. This might occur if the
local electronic structure was modified significantly and seems to be consistent with a number of other recent STS results.\cite{nunner:2006,andersen:2006,andersen2:2006,andersen:2007,slezak:2008}

Nunner {\it et al}.\cite{nunner:2005} observed on phenomenological grounds  that this simple
assumption of dopant atoms modulating the pair interaction could
explain  a large number of observed
correlations; for example, it correctly reproduces the anti-correlation
of coherence peak height and peak energy, the correlation of impurity
position with gap size, and the detailed frequency dependence of
the O:LDOS($\omega$) correlation.   In almost all treatments of disorder in
superconductors, impurities are assumed to simply scatter
electrons via a screened Coulomb potential, rather than modify
the pair interaction locally.  The well-known suppression of the gap due to pairbreaking
is of course observed in such theories if they treat the gap self-consistently; this effect is distinct, however, from a modulation of the underlying interaction, which may be expected to influence the gap in a potentially new way.
Exceptions to the standard approach in the literature on conventional superconductors include Refs. \onlinecite{Earlytau1}, where the notion of a local pair interaction was considered in some specialized contexts; from these works it is clear that these effects will be largest in systems  where the coherence length is small, like the cuprates.

The goal of such theories has been to use the STS data to potentially deduce
which changes in electronic structure influence the pairing interaction locally, and thereby learn about the origin of pairing itself in these systems.  For such a program to be successful, one needs an understanding of the location of the dopants and the changes in electronic structure which they cause, as well as a microscopic model capable of including these changes.  Thus far some work has been performed on the first question.  For example,  He {\it et al.}\cite{YHe:2006} performed density functional theory (DFT) supercell
calculations to find the position of lowest energy for an O dopant in a \BSCCO  unit cell, and noted that the interstitial atom caused a rigid tilting of the CuO$_4$ half-octahedron, as well as giving rise to an additional weakly hybridized state at -1eV which coupled well to the STM tip.

In addition, considerable model-based work has been done on the second question, that of determining how an impurity might actually affect the pairing interaction.  In the strong coupling limit, superconductivity is sometimes treated within slave boson mean field theories of the
$t-J$ model,\cite{Ruckenstein,Baskaran} whereby the pairing interaction
is provided by the super-exchange interaction $J_{ij}$ itself.  Assuming some random distribution of exchange couplings,\cite{Wang,JXZhu,WChen} results were obtained for the gap maps and other observables which were similar to those found by Nunner {\it et al.}\cite{nunner:2005}  The study of the microscopic origin of pairing interaction modulations was begun using the strong coupling approach by M\'{a}ska {\it et al.},\cite{Maska} who assumed that the presence
of the dopant atoms induces a position-dependent shift of the
atomic levels in the CuO$_2$ plane, and then calculated the
locally modified super-exchange interaction $J$ between copper ions in the
presence of an impurity from a perturbation expansion of the one-band
Hubbard Hamiltonian up to second-order. Within this framework,
 the diagonal disorder in the plane always leads to an enhancement of $J$;  accordingly, with
the assumption that pairing is due to super-exchange, the
pairing superconducting gap always increases in the vicinity of the dopant atoms,
as assumed phenomenologically in Ref. \onlinecite{nunner:2005}.  However,
Johnston {\it et al.}\cite{Johnston} argued based on cluster model
calculations for the three-band Hubbard Hamiltonian,\cite{Emery}
which account explicitly for the Cu-O hopping processes,
 that electrostatic modifications due to the presence of
the oxygen dopants locally suppress $J$. An enhancement of $J$ was only possible if they assumed that electronic coupling to local phonon modes was
strongly modified by the dopants.  Foyevstova {\it et al.}\cite{Foyevstova1} returned to the perturbation theory approach in the three-band model, and showed explicitly that whether an impurity enhances or suppresses the local exchange in this somewhat more realistic case depends explicitly on details of the
Cu-O model parameters.

For completeness it should be mentioned that a recent calculation within the Gutzwiller approximation to the $t-J$ model including disorder also studied the local disorder-induced variations in the effective super-exchange couplings $J$.\cite{christensen:2011} This study, however, focused on the consequences of charge reorganization and the doping dependence of disorder-induced magnetism, and did not include the explicit $J$-enhancements caused by renormalized local Coulomb interactions.\cite{Maska}

Assuming, as in the strong coupling case, that the mean field interaction  can be taken as a local pairing strength is intuitively reasonable, but the approximation is
even  less well controlled than in the homogeneous case.
It should be possible to study this problem from the weak coupling side as well, in an approach where the effective interaction in the particle-particle channel can be clearly defined, and which  has the additional advantage that it can be more easily coupled to {\it ab initio} results which provide the needed structural and electronic structural details for a given system.  An ``average" version of the  calculation envisaged was performed by Foyevstova {\it et al.},\cite{Foyevstova2} who calculated the total singlet spin-fluctuation interaction\cite{berk:1966} for a \BSCCO~system with and without an O dopant;
in this case the defect was indeed found to enhance the overall pairing.

 Nevertheless, one would like to understand the influence of an impurity on pairing at the local level.  In this work we continue this program from the microscopic side by formulating the full RPA particle-particle vertex in the singlet channel in real space for a general inhomogeneous system.  This interaction has a somewhat more complicated mathematical form compared to the popular Berk-Schrieffer fluctuation exchange expression.\cite{berk:1966,scalapino:1995} We then evaluate the real  pairing interaction  for a single impurity in a Hubbard model, and show that remarkably large enhancements of pairing strength can be obtained near a dopant impurity. The spin-fluctuation model produces a
 characteristic shape of the pairing modulation near a strong impurity which is reflected in the LDOS pattern observable by STM.  In our conclusions we discuss the prospects for observing this behavior.

\section{Model and Method}
In the initial step of the calculation of the effective real space pairing potential, we obtain the electronic densities calculated self-consistently in the normal state using a mean-field approximation to the one-band Hubbard model
\begin{eqnarray}
H_0&=&\sum_{i,j, \sigma}t_{i,j}c_{i\sigma}^{\dagger}c_{j\sigma}+ \sum_{i \sigma} (U \langle n_{i\sigma} \rangle - \mu ) n_{i\bar\sigma} \nonumber \\
&&+ \sum_{i \sigma} V_{\rm imp} \delta(r_i-r_{\rm imp}) n_{i\sigma}.
\label{eq:H}
\end{eqnarray}
Here, $c_{i\sigma}^{\dagger}$ refers to creation of an electron with spin $\sigma$ at lattice site $i$, and $n_{i\sigma}$ is the number operator of spin $\sigma$ particles at site $i$. The hoppings $t_{i,j}$ are included to nearest $t=1$, and next-nearest neighbor sites $t'=-0.3$.
Note that this Hamiltonian also contains the impurity potential $ V_{\rm imp}$ at a site placed at position $r_{i_{\rm imp}}$. A diagonalization of Eq.(\ref{eq:H}) allows us to obtain the effective interaction $V_{\text{eff}}(i,j)$ which in real space can be written as
\begin{equation}
 V_{\text{eff}}(i,j)=U+\frac{U^3\chi^{\down\down}\chi^{\up\up}}{\hat{1}-U^2\chi^{\down\down}\chi^{\up\up}}\Big|_{(i,j)}+\frac{U^2\chi^{\down\up}}{\hat{1}-U\chi^{\down\up}}\Big|_{(i,j)}.
\label{eq:Veff}
\end{equation}
The effective interaction (\ref{eq:Veff}) is due to longitudinal and transverse spin-fluctuations using the approach of Berk and Schrieffer,\cite{berk:1966} and the
 derivation  is given in the Appendix.
The susceptibilities entering Eq.(\ref{eq:Veff}) are real space matrices and given by
\begin{eqnarray}
\chi_{ij}^{\sigma \sigma'}\!=\!\sum_{m,n}u_{mi\sigma}u_{mj\sigma}u_{nj\sigma'}u_{ni\sigma'} \frac{f(E_{m\sigma})-f(E_{n\sigma'})}{E_{n\sigma'}-E_{m\sigma}+i\eta}, \nonumber \\
\end{eqnarray}
in terms of the eigenvectors $u_{m\sigma}$ and eigenvalues $E_{m\sigma}$ obtained in the diagonalization of Eq.(\ref{eq:H}).

It is useful now to consider the partial Fourier transform of the susceptibility with respect to
relative coordinate ${\bf r}={\bf r}_i-{\bf r}_j$, leaving explicit dependence on one spatial variable, ${\bf r}_i$
\begin{equation}\label{eq:chiFT}
    \chi({\bf q},{\bf r}_i)=\sum_{{\bf r}_j} e^{i{\bf q}\cdot({\bf r}_i-{\bf r}_j)} \chi({\bf r}_i,{\bf r}_j).
\end{equation}

Note that Eq.~(\ref{eq:Veff}) reduces to the usual form in
 the case of translational invariance, $\chi({\bf q,r}_i)=\chi({\bf q})$. In addition, it is worth
 remarking that the convention used here
 is slightly different from other works, e.g. Ref. \onlinecite{ohashi} where the mixed susceptibility is a function of the
 center of mass variable $({\bf r}_i+{\bf r}_j)/2$.

After the calculation of the effective spin-fluctuation mediated pairing, the densities $\langle n_{i\sigma} \rangle$ and superconducting gap values $\Delta_{i,j}$ are calculated self-consistently from Eq.(\ref{eq:H}) with the addition of a standard BCS singlet pairing term, i.e. from the Hamiltonian
\begin{eqnarray}
H_{\rm SC}&=&H_0+\sum_{i,j}\left[\frac{\Delta_{i,j}}{2}(c_{i\up}^\dagger c_{j\down}^\dagger-c_{i\down}^\dagger c_{j\up}^\dagger)+H.c.\right],
\label{eq:Hdelta}
\end{eqnarray}
where $\Delta_{i,j}=\frac{V_{\text{eff}}(i,j)}{2}\langle c_{j\down} c_{i\up}-c_{j\up} c_{i\down} \rangle$. The factors of $1/2$ arises from the restriction to the singlet pairing channel, cf. the Appendix.

\section{Results}

\subsection{Point-like nonmagnetic impurity}

\begin{figure}[b]
 \begin{center}
 \includegraphics[width=8.5cm]{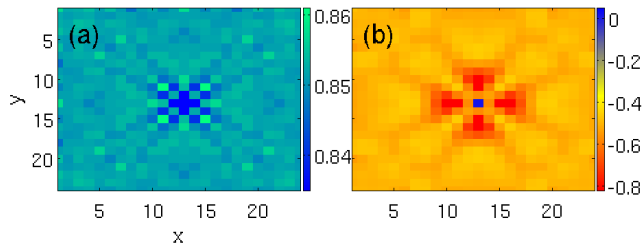}
 \end{center}
\caption{(a) Local charge density $n_i$ in the normal phase prior to calculation of the effective pairing interaction. For the results presented here the parameters are: $U=2.2$, $t'=-0.3$ (all energies are given in units of $t$) and doping $x=0.15$. A strong point-like nonmagnetic impurity ($V_{\rm imp}=10$) is situated at site $(x_{\rm imp},y_{\rm imp})=(13,13)$. (b) Effective pairing interaction between nearest neighbors for the same system as in (a). For each site the average potential to the four nearest neighbors is plotted.}
\label{FIG1}
\end{figure}

\begin{figure}[t]
 \begin{center}
 \includegraphics[width=8.cm]{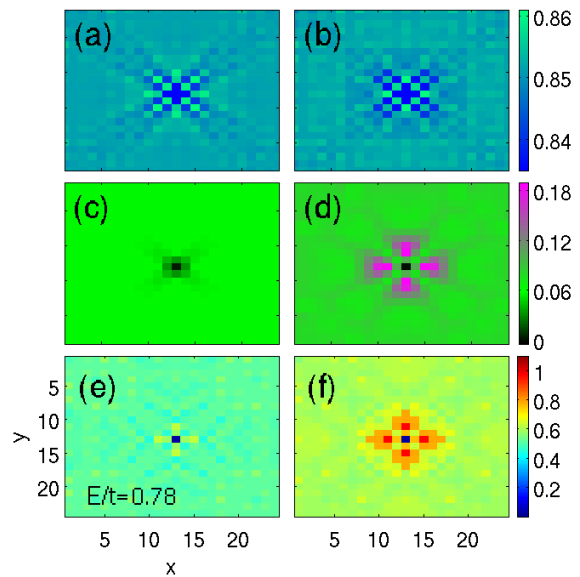}
\end{center}
 \caption{(a) Local charge density $n_i$ in a system with a strong nonmagnetic impurity ($V_{\rm imp}=10$) and a spatially homogeneous superconducting pairing potential $V_{\text{eff}}=-0.43$. Parameters: $U=2.2$, $t'=-0.3$ and doping $x=0.15$. (b) Same as (a) but in the case of the spatially inhomogeneous spin-fluctuation mediated pairing potential $V_{\text{eff}}(i,j)$ from Fig.~\ref{FIG1}(b). (c,d) Local superconducting $d$-wave order parameter corresponding to the two cases in (a) and (b), respectively. (e,f) Local density of states at energy $E=0.78$ (close to maximum gap value) corresponding to the two cases in (a) and (b), respectively.}
\label{FIG2}
\end{figure}

In Fig.~\ref{FIG1}(a) we show the normal state electron density, $n_i=\langle n_{i\up}+n_{i\down} \rangle$ as a function of site $i$ around a single nonmagnetic point-like impurity. In contrast to the superconducting case, in the normal state there is in this model never any induced local magnetization due to the absence of the impurity bound state near the Fermi level.\cite{harter:2007,andersenU:2007,andersenU:2008,andersen:2010} The effective pairing interaction between the nearest neighbor sites is shown in Fig.~\ref{FIG1}(b), where we at each site show the average interaction to the four nearest neighbors:
 $V_{\text{sc}}(i)=\frac{1}{4}\sum_{j} V_{\text{eff}}(i,j)$, $j\in \{i\pm \hat{x}, i\pm \hat{y}\}$. It is seen that the effective pairing interaction is suppressed at the impurity site, but enhanced on a quartet of small regions close to the impurity.
 The response of the system to
a perturbation is dominated by the peaks in the homogeneous susceptibility,
although because the perturbation is space dependent all wave vectors
are coupled. In the present band structure, $\chi^{\text{RPA}}({\bf q},\omega)=\chi^{0}({\bf q},\omega)/(1-U\chi^{0}({\bf q},\omega))$ at $\omega=0$ is peaked at incommensurate
wave vectors ${\bf Q} + \Delta {\bf q}$, where ${\bf Q}=(\pi,\pi)$ and $\Delta {\bf q}\simeq (\pm 0.5\pi,0), (0,\pm 0.5\pi)$. The spatial modulation of the {\it nearest neighbor} pairing shown in Fig.~\ref{FIG1}(b) is determined by  $\Delta {\bf q}$, giving rise to the quartet of peaks at roughly four lattice spacings from
the impurity site.
  With the effective pairing interaction given in Eq.(\ref{eq:Veff}), we can calculate self-consistently the densities $\langle n_{i\sigma} \rangle$ and the superconducting $d$-wave order parameter $\Delta_{i,j}$ from $H_{\rm SC}$, see Eq.(\ref{eq:Hdelta}). For all results shown in this paper we include the twelve
 closest sites $j$ around each site $i$ in $V_{\text{eff}}(i,j)$. The resulting spatial structure of the density modulations and the superconducting gap variations are shown in Fig.~\ref{FIG2}(b,d). The superconducting order parameter is averaged over the four nearest neighbor sites $\Delta_i=\frac{1}{4}[ \Delta_{i,i+\hat{x}}+\Delta_{i,i-\hat{x}}-\Delta_{i,i+\hat{y}}-\Delta_{i,i-\hat{y}}]$.
For comparison, we also calculated the densities and the $d$-wave order parameter in a system containing the same impurity, but with a homogeneous pairing potential calculated from a clean system with the same parameters, giving a nearest neighbor pairing interaction $V_{\text{sc}}=-0.43$. The result is shown in Fig.~\ref{FIG2}(c), where it is seen that no enhancement of the $d$-wave gap is present, as opposed to the case of spin-fluctuation mediated pairing (Fig. \ref{FIG2}(d)), where the $d$-wave gap is locally enhanced and the spatial structure is similar to the structure of the effective pairing potential from Fig.~\ref{FIG1}(b). Note that the scales of the pairing interaction $V_{\text{eff}}$
and superconducting gap $\Delta$ modulations are similar.  In a clean superconductor, the
order parameter response to a local perturbation will be smeared out over a coherence length
$\xi_0$, which in a realistic system is larger than the atomic lattice constant $a$; here, however we have shown results for an artificially large gap, corresponding to a coherence length $\xi_0\simeq a$, in order to display the
desired physical effects more clearly.

The LDOS is suppressed at the impurity site both in the case of a constant effective interaction and in the case of spin-fluctuation mediated pairing, see Fig.~\ref{FIG2}(e,f). However, whereas the LDOS in the case of constant effective interaction is roughly constant for all sites away from the impurity the picture is different when the pairing interaction is due to spin fluctuations. In the latter case, we find significant LDOS enhancements at energies close to the gap value at the same quartet of real space  regions seen in Figs.~\ref{FIG1}(b) and \ref{FIG2}(d).

\begin{figure}[t!]
 \includegraphics[width=7.5cm]{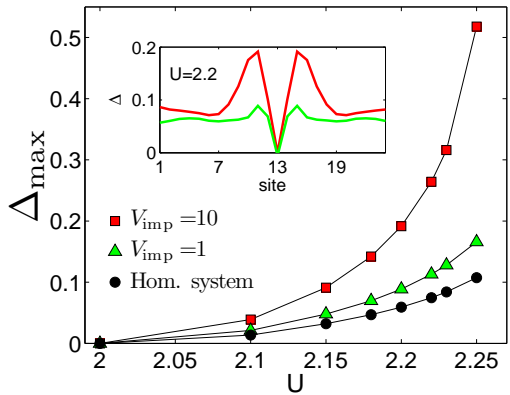}
 \caption{Maximum of the superconducting $d$-wave gap, $\Delta_{\rm max}$, as a function of Coulomb interaction $U$ for point-like nonmagnetic impurities and a homogeneous system. For all runs the system displays no antiferromagnetic order, since the critical $U$ in this system is $U_{c2}>2.4$. The inset shows the profile of the $d$-wave gap as a function of site $(x,y_{\rm imp})$}
\label{FIG3}
\end{figure}

\begin{figure}[t!]
 \begin{center}
\includegraphics[width=8cm]{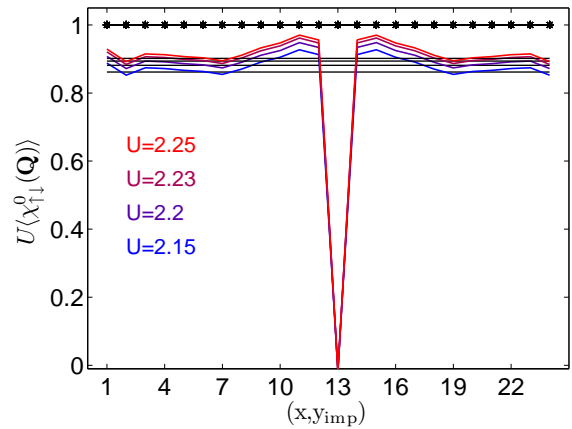}
 \caption{Average of $U \chi_{\up \down}^0({\bf Q})$ around ${\bf Q}=(\pi,\pi)$ for different values of $U$ as a function of site ${\bf r}_i=(x,y_{\rm imp})$. Black lines are $U \langle \chi_{\up \down}^0({\bf Q})\rangle$ for a homogeneous system whereas colored lines are for systems with a nonmagnetic impurity at $(x_{\rm imp},y_{\rm imp})=(13,13)$.}
\label{FIG4}
\end{center}
\end{figure}

From Fig.~\ref{FIG2}(d) it is clear that certain sites close to the impurity site exhibit a local maximum of the superconducting gap.
In Fig.~\ref{FIG3} we show how this maximum gap value depend on both the strength of the Coulomb interaction $U$
 and the impurity potential $V_{\rm imp}$. For reference, we have plotted the gap in a homogeneous system which is also enhanced upon increased $U$. This effect arises simply from the singularity in the RPA susceptibility. Since we tune the system towards the singularity from below in the paramagnetic phase, it is the last term in the effective interaction of Eq.(\ref{eq:Veff}) which is the most important.
In the system containing a single impurity, the gap is strongly enhanced as was shown in Fig.~\ref{FIG3}. The local structure of the gap enhancement is similar for a strong and a weak impurity as seen from the inset in Fig.~\ref{FIG3} which shows a profile of the $d$-wave gap through the impurity site. However, the enhancement is much more pronounced in the strong impurity case. \\
The local enhancement effect can be understood from spatially-dependent variations in the bare transverse spin susceptibility. Due to the finite value of next-nearest neighbor interaction, $t^\prime=-0.3$, the susceptibility displays splitting in reciprocal space into incommensurate peaks and a broad maximum around the ordering vector ${\bf Q}= (\pi,\pi)$. We Fourier transform the real space susceptibility to obtain $\chi_{\up \down}({\bf q,r}_i)$ and map out the change in the magnitude of the susceptibility near ${\bf Q}$ as a function of site ${\bf r}_i$. To account for the broadening of the peak and to minimize finite-size effects, we average over a region containing ${\bf Q}$ and the eight closest $q$-values around ${\bf Q}$ to obtain $\langle \chi_{\up \down}^0 ({\bf Q,r}_i)\rangle $. The result of this procedure is given in Fig.~\ref{FIG4}.
It is seen that due to local variations in the bare transverse spin susceptibility, the quartet of points with enhanced pairing in Figs.~\ref{FIG1}(b) and \ref{FIG2}(d,f) are characterized by being locally closer to the Stoner instability $U\chi^0_{\up\down}({\bf Q})\rightarrow 1$ than any other sites in the system. This is the origin of the observed enhancement of the pairing potential and ultimately what causes the local gap enhancements.
Specifically, charge density variations due to an impurity act to tune the system closer to the Stoner instability locally which results in gap enhancements in the neighboring regions of the impurity.\\

\begin{figure}[h!]
\begin{center}
\includegraphics[width=7.5cm]{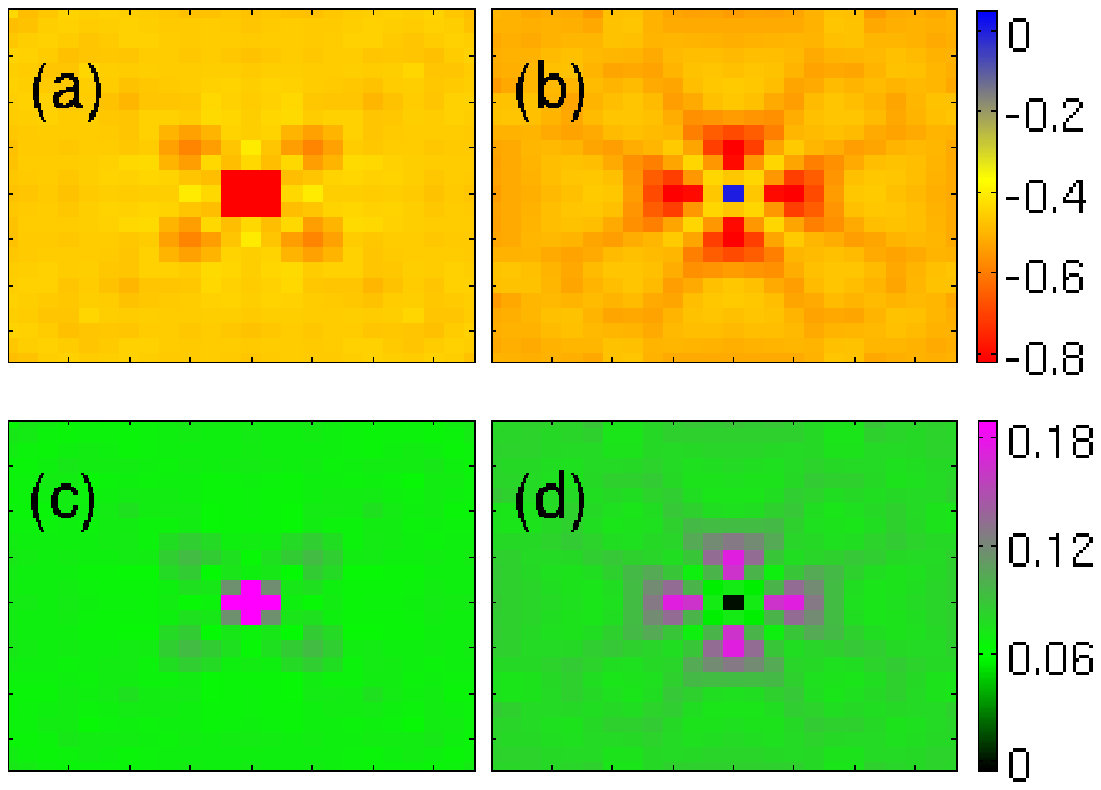}
\includegraphics[width=7.5cm]{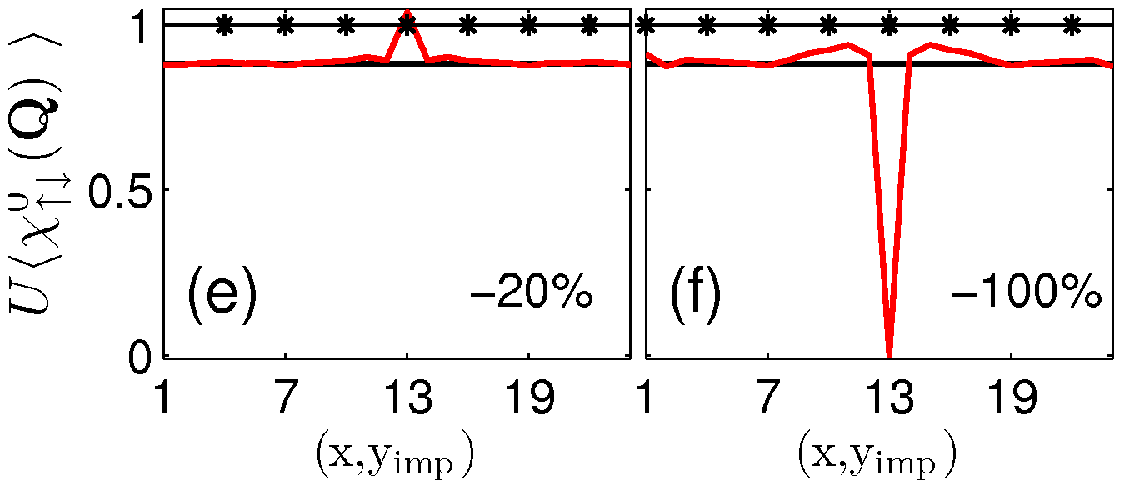}
  \end{center}
 \caption{(a,b) Effective pairing interaction between nearest neighbors. For the results presented here the parameters are: $U=2.2$, $t'=-0.3$, doping $x=0.15$. A reduced hopping impurity of strength $-20$ \% (a) and $-100$ \% (b) is situated at site $(x_{\rm imp},y_{\rm imp})=(13,13)$. (c,d) Local superconducting $d$-wave order parameter for the two impurity cases. (e,f) Average of $U \chi_{\up \down}^0({\bf Q})$ around ${\bf Q}=(\pi,\pi)$ as a function of site ${\bf r}_i=(x,y_{\rm imp})$.
}
\label{FIG5}
\end{figure}

\subsection{Impurity with reduced hopping constants}

In this section we present the results of the local pairing modulations for another kind of impurity characterized by all hopping constants onto the impurity site being reduced by a certain percentage. In the extreme limit of $100$\% reduction, we obtain results similar to the case of a strong nonmagnetic potential scatterer as expected  (compare e.g. Fig.~\ref{FIG2}(d) and Fig.~\ref{FIG5}(d)). Both magnitude and local structure of the superconducting $d$-wave gap is similar in these two cases. Furthermore, there is no induced local antiferromagnetic order. In the intermediate regime, however, where the reduction is less than $100$\%, we find a different local structure of the $d$-wave gap enhancement. As an example, we show in Fig.~\ref{FIG5}(a,c) the effect of reducing the hopping elements by $20$\%. In this case weak local antiferromagnetism is induced (not shown), and the enhancement of the local pairing interaction as well as the $d$-wave gap is maximal at the impurity site. Again this gap structure can be understood from the local modulations in  the transverse spin susceptibility, see Fig.~\ref{FIG5}(e). Note that for the effective pairing interaction shown in Fig.~\ref{FIG5}(a), the spatial structure of the effective interaction and  the size of the amplitude modulations are closely comparable to the phenomenological values used in Ref. \onlinecite{nunner:2005} to model the STS gap variations in optimally doped \BSCCO.

\section{Conclusions}

The results of the local gap enhancements found in this paper gives rise to a number of additional questions for future studies. First, is the LDOS within the present formalism from more realistic many-impurity simulations consistent with the experimental STS data? As shown by Nunner {\it et al.}\cite{nunner:2005} the positive correlation between dopant atoms and large gap regions will certainly exist with the presented scenario, but the LDOS evolution with doping, where pseudogap regions become dominant in the underdoped regime cannot be captured within this weak-coupling approach. In addition, several unsolved LDOS issues remain regarding the origin of the so-called extinction of impurity resonances near strong onsite impurity potentials in large-gap regions.\cite{andersen:2008}

Second, what are the effects on a modified pairing near other kinds of impurities? We have recently applied the present formalism to also study the pairing interaction and local gap structure near point-like magnetic impurities.\cite{roemer:2012} There it was found that the local structure of the gap enhancement is very similar compared to the case of point-like nonmagnetic impurities even though local antiferromagnetism is induced around a magnetic impurity site.

Third, what is the feedback effect on the pairing interaction in coexisting phases where e.g. spin- and charge density waves are present? The current formulation of the real space pairing mechanism
does not hold in the ordered state. The simple expression that was derived in Eq.(\ref{eq:Veff}) predicts a singularity in the effective pairing interaction as the system develops long range antiferromagnetic order, as has been discussed previously. This effective vertex is, however,
known to be suppressed by interactions with spin waves (Adler principle\cite{Schrieffer95,Chubukov02}), and our model is therefore only valid in the paramagnetic phase.

Fourth, what are the thermodynamic consequences of the local enhancements of the gap found above? In particular one might expect a non-trivial evolution of the residual resistivity $\Delta \rho$ and $T_c$ suppression with increasing disorder concentrations.  {The slow rate of $T_c$ suppression relative to $\Delta \rho$ in the case of
strong scatterers such as Zn in cuprates has never been satisfactorily explained, but has been attributed to
correlation effects.\cite{siggi2007}} This interesting question will be the topic of a future study.

Finally, how robust is the present result of local {\it enhancement}
of pairing to the form of the bare interaction vertex?  In the strong coupling limit, Foyevstova {\it et al.}\cite{Foyevstova1}
showed that the local pairing interaction (exchange) was increased only in the limit of
pure Hubbard-type correlations in a single band.\cite{Maska}  We have established here the weak
coupling analog of this result, but it is clear that other types of interactions in the host, beyond the scope of this work,  might lead to different results.

Summarizing, we have studied how single nonmagnetic potential and hopping impurities modify the local pairing interaction in a $d$-wave superconductor. The local pairing interaction was calculated within a real space formulation of the spin-fluctuation mechanism, and this formalism was used to study the single-impurity problem. We find that the local variations in the charge densities or hopping integrals may significantly enhance the local $d$-wave gap close to the impurity sites due to a locally enhanced transverse spin susceptibility. We have discussed the observation of such enhancement effects in relation to STS measurements on the surface of cuprate superconductors. Our model provides a powerful tool to study ideas about the effect of inhomogeneity on superconductivity
at a microscopic level.  It will be interesting to find whether inhomogeneous defects or magnetic structures can
enhance the critical temperature, as was found in purely phenomenological approaches.\cite{martin_etal}

\section{Acknowledgements}

B.M.A. acknowledges support from The Danish Council for Independent Research $|$ Natural Sciences.  P.J.H. acknowledges support from NSF-DMR-1005625.

\begin{appendix}
 \section{Effective pairing interaction in real space}
\begin{figure}[t!]
\includegraphics[width=8cm]{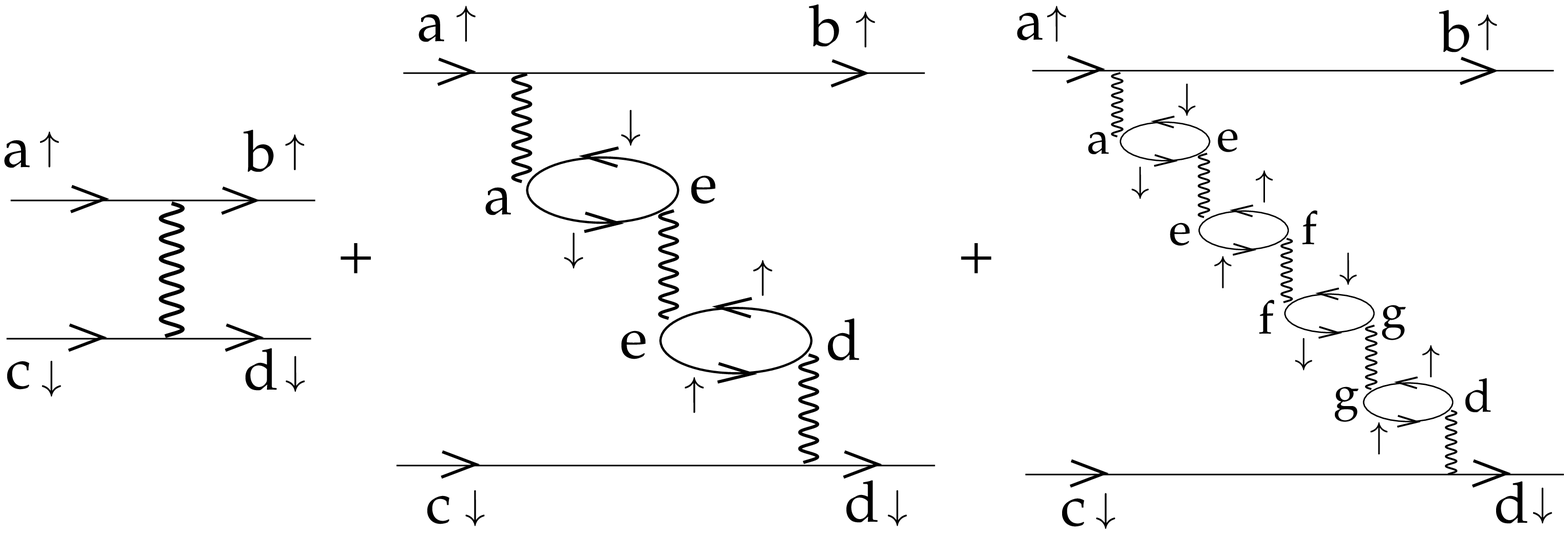}
 \caption{Real space diagrams for longitudinal spin fluctuations to fifth order in $U$.}
\label{fig:bubble}
\end{figure}

\begin{figure}
\includegraphics[width=8cm]{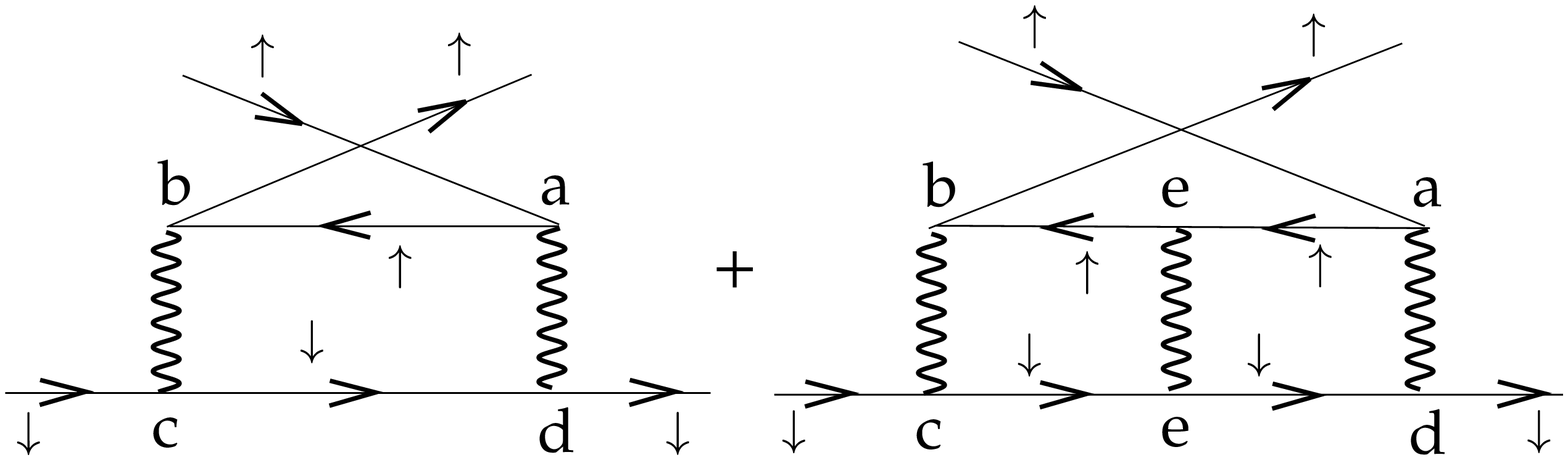}
 \caption{Real space diagrams for transverse spin fluctuations to third order.}
\label{fig:ladder}
\end{figure}

The interaction Hamiltonian is given by
\begin{equation}
 H_{\rm int}=\sum_{a,b,c,d}V_{\rm eff}(a,c;b,d)\big(c_{b \up}^\dagger c_{d \down}^\dagger c_{c \down} c_{a \up} + H.c.  \big),
\end{equation}
where $a,..d$ are real space points.
We define the noninteracting spin-dependent susceptibility of the normal phase as
\begin{equation}
 \chi_{ab}^{\sigma \sigma'}(\tau)=-\mathcal{G}_{ab,\sigma}(\tau)\mathcal{G}_{ba,\sigma'}(-\tau).
\end{equation}
Note the sign convention. The effective interaction in real space is evaluated directly from the real space version of the standard spin-fluctuation diagrams. Since the Coulomb interaction is local both in space and time, it connects fermions at the same site and of opposite spin. Omitting the external fermion lines and evaluating the interaction part from the bubble diagrams shown in Fig.~\ref{fig:bubble} gives
\begin{align}
 V_{\rm eff}^{(1)}(a,c;b,d)=
U\delta_{ab} \delta_{cd}\delta_{ad} +\nonumber \\
U^3\sum_{e}\delta_{ab}\delta_{cd}\mathcal{G}_{a e,\down}\mathcal{G}_{ea,\down}\mathcal{G}_{ed,\up}\mathcal{G}_{de,\up}+\cdots
\end{align}
The pairing interaction is dependent only on two spatial variables since $U$ is onsite. We get
\begin{align}
V_{\rm eff}^{(1)}(a,c;b,d)=\delta_{ab}\delta_{cd}[ U \delta_{ad} +U^3\sum_{e}\chi_{ae}^{\down\down}\chi_{ed}^{\up\up}\nonumber \\
+ U^5\sum_{e,f,g}\chi_{ae}^{\down\down}\chi_{ef}^{\up\up}\chi_{fg}^{\down\down}\chi_{gd}^{\up\up}+\cdots]\nonumber \\
=\delta_{ab}\delta_{cd}[U\hat{1}+U^3\chi^{\down\down}\chi^{\up\up}+U^5(\chi^{\down\down}\chi^{\up\up})^2+\cdots]_{ad}\nonumber\\
=\delta_{ab}\delta_{cd} \left( U\hat{1}_{ad}+\frac{U^3\chi^{\down\down}\chi^{\up\up}}{\hat{1}-U^2\chi^{\down\down}\chi^{\up\up}}\Big|_{ad}\right).
\label{eq:V1}
\end{align}
Similarly, in the case of the ladder diagrams shown in Fig.~\ref{fig:ladder} we get
\begin{align}
 V_{\rm eff}^{(2)}(a,c;b,d)=\delta_{ad}\delta_{bc}[
U\chi_{ab}^{\up\down} +U^3\sum_{e}\chi_{ae}^{\up\down}\chi_{eb}^{\up\down}\nonumber \\
+U^5 \sum_{e,f}\chi_{ae}^{\up\down}\chi_{ef}^{\up\down}\chi_{fb}^{\up\down}+\cdots] \nonumber \\
=\delta_{ad}\delta_{bc}[U^2\chi^{\up\down}+U^3{\chi^{\up\down}}^2+U^5{\chi^{\up\down}}^4+\cdots]\nonumber\\
=\delta_{ad}\delta_{bc}\left(\frac{U^2\chi^{\up\down}}{\hat{1}-U\chi^{\up\down}}\Big|_{ab}\right).
\label{eq:V2}
\end{align}
The interaction Hamiltonian now becomes
\begin{align}
 H_{\rm int}=\sum_{a,b,c,d}V_{\rm eff}^{(1)}(a,c;b,d)\big(c_{b \up}^\dagger c_{d \down}^\dagger c_{c \down} c_{a \up} + H.c.  \big) \nonumber \\
\hspace{0.7cm}+V_{\rm eff}^{(2)}(a,c;b,d)\big(c_{b \up}^\dagger c_{d \down}^\dagger c_{c \down} c_{a \up} + H.c.  \big)\nonumber\\
=\sum_{ad}V_{\rm eff}^{(1)}(a,d;a,d)\left( c_{a \up}^\dagger c_{d \down}^\dagger c_{d \down} c_{a \up} + H.c.  \right)\nonumber \\
+\sum_{ab}V_{\rm eff}^{(2)}(a,b;b,a)\left( c_{b \up}^\dagger c_{a \down}^\dagger c_{b \down} c_{a \up} + H.c.  \right).
\label{Hint}
\end{align}
From this it is apparent that the first part corresponds to spin-conserving processes while the second describes a spin-flip interaction.

\subsection{Mean field Hamiltonian}
The result for the effective interaction derived in Eq.~(\ref{Hint}) is treated in a mean-field approach. We consider only the singlet channel.
\begin{eqnarray}
 H_{\rm int}&=&2\sum_{i,j}V_{\rm eff}^{(1)}c_{i \up}^\dagger c_{j \down}^\dagger c_{j \down} c_{i \up}\nonumber \\
&&-\sum_{i,j}V_{\rm eff}^{(2)}(c_{i \down}^\dagger c_{j \up}^\dagger c_{j \down} c_{i \up}+c_{i \up}^\dagger c_{j \down}^\dagger c_{j \up} c_{i \down})\nonumber \\
 &=&H_{\rm int}^{(1)}+ H_{\rm int}^{(2)}
\end{eqnarray}
where we have used that the interactions are symmetric with respect to interchange of spatial indices.
We define two gaps which are respectively symmetric (s) and anti-symmetric (a) under interchange of spatial indices
\begin{eqnarray}
 \Delta_{(1),ij}^{s/a}&=&\frac{V_{\rm eff}^{(1)}}{2}\left(\langle c_{j\down}c_{i\up}\rangle \pm \langle c_{j\up}c_{i\down}\rangle \right), \nonumber \\
 \Delta_{(2),ij}^{s/a}&=&\frac{V_{\rm eff}^{(2)}}{2}\left(\langle c_{j\down}c_{i\up}\rangle \pm \langle c_{j\up}c_{i\down}\rangle \right).
\end{eqnarray}
For the first term in the interaction Hamiltonian, we get by a standard mean field decoupling
\begin{align}
 H_{\rm int}^{(1)}=2\sum_{i,j}V_{\rm eff}^{(1)}c_{i \up}^\dagger c_{j \down}^\dagger c_{j \down} c_{i \up} \nonumber \\
\simeq \sum_{i,j}V_{\rm eff}^{(1)}\left( \langle c_{i \up}^\dagger c_{j \down}^\dagger \rangle c_{j \down} c_{i \up} + c_{i \up}^\dagger c_{j \down}^\dagger \langle c_{j \down} c_{i \up} \rangle\right)\nonumber \\
=\sum_{ij}\frac{1}{2}\Big[\Delta_{(1),ij}^{s^*}(c_{j\down}c_{i\up} -c_{j\up}c_{i\down})+
\Delta_{(1),ij}^{s}(c_{i\up}^\dagger c_{j\down}^\dagger -c_{i\down}^\dagger c_{j\up}^\dagger)\nonumber \\
\hspace{0.7cm}+
\Delta_{(1),ij}^{a^*}(c_{j\down}c_{i\up} +c_{j\up}c_{i\down})+
\Delta_{(1),ij}^{a}(c_{i\up}^\dagger c_{j\down}^\dagger +c_{i\down}^\dagger c_{j\up}^\dagger)
\Big].\nonumber
\end{align}
The second term gives
\begin{align}
 H_{\rm int}^{(2)}=-\sum_{i,j}V_{\rm eff}^{(2)}(c_{i \down}^\dagger c_{j \up}^\dagger c_{j \down} c_{i \up}
+c_{i \up}^\dagger c_{j \down}^\dagger c_{j \up} c_{i \down})\nonumber \\
=-\sum_{i,j} \frac{V_{\rm eff}^{(2)}}{2}\Big[
\langle c_{i \down}^\dagger c_{j \up}^\dagger \rangle c_{j \down} c_{i \up}
+\langle c_{i \up}^\dagger c_{j \down}^\dagger \rangle c_{j \up} c_{i \down} \nonumber \\
 \hspace{1.7cm}
+ c_{i \down}^\dagger c_{j \up}^\dagger  \langle c_{j \down} c_{i \up}\rangle
+ c_{i \up}^\dagger c_{j \down}^\dagger  \langle c_{j \up} c_{i \down} \rangle \Big] \nonumber \\
=
\sum_{ij}\frac{1}{2}\Big[\Delta_{(2),ij}^{s^*}(c_{j\down}c_{i\up} -c_{j\up}c_{i\down})+
\Delta_{(2),ij}^{s}(c_{i\up}^\dagger c_{j\down}^\dagger -c_{i\down}^\dagger c_{j\up}^\dagger)\nonumber \\
\hspace{0.7cm}+
\Delta_{(2),ij}^{a^*}(c_{j\down}c_{i\up} +c_{j\up}c_{i\down})+
\Delta_{(2),ij}^{a}(c_{i\up}^\dagger c_{j\down}^\dagger +c_{i\down}^\dagger c_{j\up}^\dagger)
\Big].\nonumber
\end{align}
Adding these results, we obtain the final effective mean field Hamiltonian in the singlet channel
\begin{equation}
 H_{\rm  singlet}^{\rm MF}=
\frac{1}{2}\sum_{i,j}\big[\Delta_{ij}^{s*}(c_{j\downarrow}c_{i\uparrow}-c_{j\uparrow}c_{i\downarrow})+\Delta_{ij}^{s}(c_{i\uparrow}^{\dagger}c_{j\downarrow}^{\dagger}-c_{i\downarrow}^{\dagger}c_{j\uparrow}^{\dagger}) \big],
\end{equation}
 where
\begin{equation}
 \Delta_{ij}^{s}=\frac{V_{{\rm eff}}(i,j)}{2}\big(\langle c_{j\downarrow} c_{i\uparrow}\rangle-\langle c_{j\uparrow} c_{i\downarrow}\rangle\big).
\end{equation}
Here, the total effective pairing interaction $V_{{\rm eff}}(i,j)$ is the sum of the effective interactions $V_{\rm eff}^{(1)}$ and $V_{\rm eff}^{(2)}$ derived in Eqs.~(\ref{eq:V1}) and (\ref{eq:V2}). Thus, the effective pairing is given by the real space matrix
\begin{eqnarray}
V_{\rm eff}(i,j)&=&V_{{\rm eff}}^{(1)}(i,j)+V_{{\rm eff}}^{(2)}(i,j) \nonumber \\
&=&  U\hat{1}_{i,j}+\frac{U^3\chi^{\down\down}\chi^{\up\up}}{\hat{1}-U^2\chi^{\down\down}\chi^{\up\up}}\Big|_{i,j}+ \frac{U^2\chi^{\up\down}}{\hat{1}-U\chi^{\up\down}}\Big|_{i,j}, \nonumber \\
\end{eqnarray}
\label{app:Veff}
which reduces to the usual expression in the translationally invariant case.
\end{appendix}


\begin{thebibliography}{00}

\bibitem{Fischer}  \O. Fischer, M. Kugler, I. Maggio-Aprile, C. Berthod, and C. Renner, Rev. Mod. Phys. {\bf 79}, 353 (2007).
%
\bibitem{cren} T. Cren, D. Roditchev, W. Sacks, and J. Klein, Europhys. Lett. {\bf 54}, 84 (2001).
%
\bibitem{howald} C. Howald, P. Fournier, and A. Kapitulnik, Phys. Rev. B {\bf 64}, 100504(R) (2001).
%
\bibitem{davisinhom1} S. H. Pan {\it et al.},  Nature {\bf 413}, 282 (2001).
%
\bibitem{davisinhom2} K. M. Lang, V. Madhavan, J. E. Hoffman, E. W. Hudson, H. Eisaki, S. Uchida, and J. C. Davis, Nature (London) {\bf 415}, 412 (2002).
%
\bibitem{YKohsaka:2007}
Y. Kohsaka, C. Taylor, K. Fujita, A. Schmidt, C. Lupien, T. Hanaguri, M. Azuma, M. Takano, H. Eisaki, H. Takagi, S. Uchida, J. C. Davis, Science {\bf 315}, 1380-1385 (2007).
%
\bibitem{KMcElroy_05} K.
McElroy, J. Lee, J. A. Slezak, D.-H. Lee, H. Eisaki, S. Uchida, and J. C. Davis, Science {\bf 309}, 1048 (2005).
%
\bibitem{Hoffman} J. Hoffman, {\it preprint}.
%
\bibitem{nunner:2005} T. S. Nunner, B. M. Andersen, A. Melikyan, and P. J. Hirschfeld, Phys. Rev. Lett. {\bf 95}, 177003 (2005).
%
\bibitem{nunner:2007} T. S. Nunner, P. J. Hirschfeld, B. M. Andersen, A. Melikyan, and K. McElroy, Physica C {\bf 460-462}, 446 (2007).
%
\bibitem{nunner:2006} T. S. Nunner, W. Chen, B. M. Andersen, A. Melikyan, and P. J. Hirschfeld, Phys. Rev. B {\bf 73}, 104511 (2006).
%
\bibitem{andersen:2006} B. M. Andersen, A. Melikyan, T. S. Nunner, and P. J. Hirschfeld, Phys. Rev. Lett. {\bf 96}, 097004 (2006).
%
\bibitem{andersen2:2006} B. M. Andersen, A. Melikyan, T. S. Nunner, and P. J. Hirschfeld, Phys. Rev. B {\bf 74}, 060501(R) (2006).
%
\bibitem{andersen:2007} B. M. Andersen, P. J. Hirschfeld, and J. A. Slezak, Phys. Rev. B {\bf 76}, 020507(R) (2007).
%
\bibitem{slezak:2008} J. A. Slezak, Jinho Lee, M. Wang, K. McElroy, K. Fujita, B. M. Andersen, P. J. Hirschfeld, H. Eisaki, S. Uchida, and J. C. Davis, Proc. Natl. Acad. Sci. USA {\bf 105}, 3203 (2008).
%
\bibitem{Earlytau1}
H. Suhl, D. R. Fredkin, J. S. Langer, and B. T. Matthias Phys.
Rev. Lett. {\bf 9}, 63 (1962); A. I. Larkin,  Sov. Phys. JETP
{\bf 31}, 784 (1970);  A. Weinkauf and J. Zittartz, Z. Phys. B
{\bf 21}, 135 (1975);  K. Chattopadhyay, R. A. Klemm and D. Sa, J.
Phys.: Cond. Matt. {\bf 14}, L577 (2002).
%
\bibitem{YHe:2006} Y. He, T. S. Nunner, P. J. Hirschfeld, and H.-P. Cheng, Phys. Rev. Lett. {\bf 96}, 197002 (2006).
%
\bibitem{Ruckenstein} A. E. Ruckenstein, P. J. Hirschfeld, and J. Appel, Phys. Rev.  B {\bf 36}, 857 (1987).
%
\bibitem{Baskaran} G. Baskaran,  Z. Zou, and P.W. Anderson, Solid State Comm. {\bf 63}, 973 (1987).
%
\bibitem{Wang} Z. Wang, J. R. Engelbrecht, S. Wang, H. Ding, and S. H. Pan, Phys. Rev. B {\bf 65}, 064509 (2002).
%
\bibitem{JXZhu} J.-X. Zhu, cond-mat/0508646.
%
\bibitem{WChen} W. Chen, M. Gabay, and P. J. Hirschfeld, New J. Phys. {\bf 14},  033004 (2012).
%
\bibitem{Maska} M. M. Ma\'{s}ka, \'{Z}. \'{S}led\'{z}, K. Czajka, and M. Mierzejewski, Phys. Rev. Lett. {\bf 99}, 147006 (2007).
%
\bibitem{Johnston} S. Johnston, F. Vernay, B. Moritz, Z.-X. Shen, N. Nagaosa, J. Zaanen, and T. P. Devereaux, Phys. Rev. B {\bf 82}, 064513 (2010).
%
\bibitem{Emery} V. J. Emery, Phys. Rev. Lett. {\bf 58}, 2794 (1987).
%
\bibitem{Foyevstova1} K. Foyevtsova, R. Valent\'{\i}, and P. J. Hirschfeld, Phys. Rev. B {\bf 79}, 144424(2009).
%
\bibitem{christensen:2011} R. B. Christensen, P. J. Hirschfeld, and B. M. Andersen, Phys. Rev. B {\bf 84}, 184511 (2011).
%
\bibitem{Foyevstova2}  K. Foyevtsova, H. C. Kandpal, H. O. Jeschke, S. Graser, H.-P. Cheng, R.Valent\'{\i} and P. J. Hirschfeld, Phys. Rev. B {\bf 82}, 054514 (2010).
%
\bibitem{berk:1966} N. F. Berk and J. R. Schrieffer, Phys Rev. Lett. {\bf 17}, 433 (1966).
%
\bibitem{scalapino:1995} D. J. Scalapino, Phys. Rep. {\bf 250}, 329 (1995).
%
\bibitem{harter:2007} J. W. Harter, B. M. Andersen, J. Bobroff, M. Gabay, and P. J. Hirschfeld, Phys. Rev. B {\bf 75}, 054520 (2007).
%
\bibitem{andersenU:2007} B. M. Andersen, P. J. Hirschfeld, A. P. Kampf, and M. Schmid, Phys. Rev. Lett. {\bf 99}, 147002 (2007).
%
\bibitem{andersenU:2008} B. M. Andersen and P. J. Hirschfeld, Phys. Rev. Lett. {\bf 100}, 257003 (2008).
%
\bibitem{andersen:2010} B. M. Andersen, S. Graser, and P. J. Hirschfeld, Phys. Rev. Lett. {\bf 105}, 147002 (2010).
%
\bibitem{ohashi} Y. Ohashi, Phys. Rev. B {\bf 60}, 15388 (1999); Phys. Rev. B {\bf 66}, 054522 (2002).
%
\bibitem{Schrieffer95} J. R. Schrieffer, J. Low Temp. Phys., {\bf 99}, 314 (1995).
%
\bibitem{Chubukov02} A. V. Chubukov, D. Pines, and J. Schmalian, `The Physics of Conventional and Unconventional Superconductors' edited by K.H. Bennemann and J.B. Ketterson (Springer-Verlag 2004).
%
\bibitem{andersen:2008} B. M. Andersen, S. Graser, and P. J. Hirschfeld, Phys. Rev. B {\bf 78}, 134502 (2008).
%
\bibitem{roemer:2012} A. T. R\o mer, S. Graser, P. J. Hirschfeld, and B. M. Andersen, submitted to J. Supercond. Novel Magn. (2012).
%
\bibitem{siggi2007} S. Graser, P. J. Hirschfeld, L.-Y. Zhu, and T. Dahm, Rev. B {\bf 76}, 054516 (2007).
%
\bibitem{martin_etal} I. Martin, D. Podolsky, and S. A. Kivelson, Phys. Rev. B {\bf 72}, 060502 (2005).

\end{thebibliography}
\end{document}